\documentstyle[ijmpd-la]{article}

\def\leftcontract{\mathop{\hbox{\vrule height0.5pt width6pt \vrule width0.5pt 
   height6pt}}}
\def\rightcontract{\mathop{\hbox{\vrule width0.5pt height6pt%
  \vrule height0.5pt width6pt}}}

\def\dualp#1{{}^{\ast_{(\hbox{$\scriptstyle #1$})}} \kern-1pt}
\def\dual{\,{}^\ast{}\kern-1.5pt}
\def\SS{{\cal S}} \def\V{{\cal V}}

\def\undersim#1{\mathop{\vtop{\ialign{##\crcr
     $\hfil\displaystyle{#1}\hfil$\crcr\noalign
     {\kern1pt\nointerlineskip}\hbox{$\hfil{\scriptscriptstyle \sim}\hfil$}\crcr
     \noalign{\kern1pt}}}}}

\def\Lie{\hbox{\it\char'44}} \let\pounds=\Lie

\def\strut{\hbox{\vrule width 0pt height 9.5pt depth 3.5pt}}

\begin{document}
\setlength{\textheight}{7.7truein}    

\runninghead{Bini, Germani, Jantzen}{Integral formulation of Maxwell's equations}

\normalsize\textlineskip
\thispagestyle{empty}
\setcounter{page}{1}

\copyrightheading{}			

\vspace*{0.88truein}

\fpage{1}

  \centerline{\bf
GRAVITOELECTROMAGNETISM AND THE INTEGRAL
  }\vspace*{0.035truein}
  \centerline{\bf
 FORMULATION OF MAXWELL'S EQUATIONS
  }

  \vspace*{0.37truein}
  \centerline{\footnotesize 
DONATO BINI
  }
  \vspace*{0.015truein}
  \centerline{\footnotesize\it 
Istituto per Applicazioni della Matematica C.N.R., 
I--80131 Napoli, Italy 
  and
  }
  \baselineskip=10pt
  \centerline{\footnotesize\it 
International Center for Relativistic Astrophysics, 
University of Rome, I--00185 Roma, Italy
  }
  \vspace*{10pt}
  \centerline{\footnotesize 
CRISTIANO GERMANI
  }
  \vspace*{0.015truein}
  \centerline{\footnotesize\it 
International Center for Relativistic Astrophysics, 
University of Rome, I--00185 Roma, Italy
  }
  \vspace*{10pt}
  \centerline{\footnotesize 
ROBERT T. JANTZEN
  }
  \vspace*{0.015truein}
  \centerline{\footnotesize\it 
Department of Mathematical Sciences, 
Villanova University, Villanova, PA 19085, USA 
and
  }
  \baselineskip=10pt
  \centerline{\footnotesize\it 
International Center for Relativistic Astrophysics, 
University of Rome, I--00185 Roma, Italy
  }


  \vspace*{0.225truein}
  \publisher{[version: \today](received date)}{(revised date)}

  \vspace*{0.21truein}


\abstracts{
The integral formulation of Maxwell's equations expressed in terms of an arbitrary observer family in a curved spacetime is developed and used to clarify the meaning of the lines of force associated with observer-dependent electric and magnetic fields. 
}{}{}

\vspace*{10pt}

\vspace*{1pt}\textlineskip

\section{Introduction}

In 1982 Thorne and MacDonald\cite{thomac} discussed the integral formulation of  Max\-well's equations in curved spacetime  and used it in the following years to develop a number of applications to black hole physics, culminating in the ``membrane paradigm" for treating the black hole horizon\cite{thoprimac} and physics around black holes. 
Their approach to the integral version of  Maxwell's equations is based on the Ellis $1+3$ congruence description of those equations\cite{ellis,ellis98} but relies on the black hole spacetime splitting by the hypersurface orthogonal locally nonrotating observers, also called the zero angular momentum observers (ZAMOs). Here this approach is generalized to an arbitrary observer family in any curved spacetime using the framework of gravitoelectromagnetism,\cite{mfg} simplifying some of their calculations by using an elegant differential-form transport theorem to calculate the time derivative of surface integrals.
With this machinery, the meaning of the lines of force for 
observer-dependent electric and magnetic fields is clarified.

A partial step in this direction was taken by Van Bladel,\cite{vanbladel} who developed an extensive body of relativistic applications of Maxwell's equations in moving reference systems in Minkowski spacetime with an eye towards realistic engineering problems, with some limited steps towards extending them to curved spacetime, heavily influenced by the $1+3$ discussion of Landau and Lifshitz\cite{ll} and M\o ller\cite{moller} on Maxwell's equations in stationary spacetimes.
The classic example of Maxwell's equations in a rotating coordinate system in Minkowski spacetime\cite{cor} as well as static-observer measured electromagnetic fields in general stationary axisymmetric spacetimes both lead to integral Maxwell equations involving a pair of observer families. One is the general observer associated with the definitions of the electric and magnetic fields through the Lorentz force law on test particles and hence lines of force of the electric and magnetic fields, and the other is the hypersurface-forming observer associated with the 
hypersurface-submanifolds over which the integration is performed and used to define the flux integrals. When these observers are not the same,  the usual  flat-spacetime inertial-coordinate relationships between lines of force, flux and sources no longer hold.

In this article differential form language coupled with a modern version of the various integral transport theorems is used to obtain the set of generalized integral Maxwell equations. Some observations about lines of force are then made in concluding remarks. An appendix shows the relationship of the modern statement of the transport theorems to those from vector analysis.

\section{Maxwell's equations in $1+3$ form}

The Maxwell equations can be expressed covariantly in many ways. In differential form language one has 
\begin{equation}\label{eq:ME}
    dF = 0\ ,\qquad 
    d \dual F = -4\pi \dual J^\flat \ (or\ \delta F = -4\pi J^\flat\,) \ ,
\end{equation}
where $F$ is the Faraday electromagnetic 2-form
and $J$ is the current vector field, obeying the conservation law
\begin{equation}
\delta J^\flat =\dual d \dual J^\flat =0\ .
\end{equation}
Here $\delta=\dual d \dual$ is the divergence operator, $\dual$ is the duality operation on differential forms (or their associated index-raised contravariant tensors), and $\sharp$ and $\flat$ are the index raising and lowering operations taking a tensor to its fully contravariant or fully covariant form. 

The splitting of the electromagnetic 2-form $F$  by any observer family (with unit 4-velocity vector field $u$: $u\cdot u=-1$, metric signature $-+++$) gives the associated electric and magnetic vector fields
$E(u)$ and $B(u)$ as measured by those observers through the Lorentz force law on a test charge, and the relative charge $\rho(u)$ and current density vector field $J(u)$. 
This splitting procedure is a standard orthogonal decomposition of tensor fields associated with the orthogonal direct sum of each tangent space into the directions along $u$ and the complementary directions in the local rest space of $u$.
This  ``relative observer decomposition" of $F$ and its dual 2-form $\dual F$ is
\begin{eqnarray*}\label{eq:FEB}
      F &=& [u\wedge E(u)+\dualp{u}B(u)]^\flat\ ,\\
\dual F &=& [-u\wedge B(u)+\dualp{u}E(u)]^\flat\ ,
\end{eqnarray*}
while $J$ has the representation
\begin{equation}
J=\rho(u) u +J(u)\ .
\end{equation}

Here $\dualp{u}$ is the spatial duality operation over the local rest space of the observer family. For example, in an observer-adapted orthonormal frame $\{e_\alpha\}$ $(\alpha=0,1,2,3)$ adapted to $u =e_0$, these definitions are equivalent to
\begin{equation}
  E(u)^\alpha = F^\alpha{}_0\ ,\
  B(u)^\alpha = -\dual F^\alpha{}_0\ ,\
  \rho(u) = J^0\ ,\
  J(u)^\alpha = P(u)^\alpha{}_\beta  J^\beta\ ,
\end{equation}
where
\begin{equation}
P(u)^\alpha{}_\beta = \delta^\alpha{}_\beta + u^\alpha u_\beta
\end{equation}
is the spatial projection tensor.

If $\eta$ is the unit oriented volume 4-form ($\eta_{0123}=1$ in an oriented orthonormal frame) and $\eta(u)=u\leftcontract\eta$ is its left contraction with $u$ (the spatial volume 3-form),
equivalently 
$\eta(u)_{\alpha\beta\gamma} =u^\delta\eta_{\delta\alpha\beta\gamma}$, then the spatial cross product of two spatial vector fields is defined by the successive right contractions
\begin{equation}
     X\times_u Y 
     = \{[\eta(u)\rightcontract Y]\rightcontract X\}^\sharp
      = \dualp{u} [X \wedge Y] \ ,
\end{equation}
or in components
\begin{equation}
  [X\times_u Y]^\alpha 
   = \eta(u)^{\alpha}{}_{\beta\gamma} X^\beta Y^\gamma\ ,
\end{equation}
where $\eta(u)_{123}=1$ in an oriented, spatially oriented observer-adapted orthonormal frame. The spatial duality operation is then
\begin{equation}
  B(u)^\alpha = \frac12 \eta(u)^\alpha{}_{\beta\gamma} F^{\beta\gamma}\ .
\end{equation}
Similarly the spatial dot product of two spatial vector fields is
\begin{equation}
  X(u)\cdot_u Y(u) = P(u)_{\alpha\beta} X(u)^\alpha Y(u)^\beta\ .
\end{equation}

If $U$ is the 4-velocity of any test particle with charge $q$ and nonzero rest mass $m$, it has the orthogonal decomposition
\begin{equation}
   U = \gamma(U,u) [ u + \nu(U,u)]\ ,
\end{equation}
defining its relative velocity and associated gamma factor with respect to $u$.
The absolute derivative of $U$ with respect to a proper time parametrization along its world line is its 4-acceleration $a(U)=DU/d\tau_U$. 
The Lorentz force law then takes the form
\begin{equation}
    m a(U) 
    = q \gamma(U,u) [ E(u) + \nu(U,u)\times_u B(u)]\ .
\end{equation}

Using the notation $\nabla(u) = P(u)\nabla$ for the spatial projection of the covariant derivative, namely
\begin{equation}
[\nabla(u)_X Y ]^\alpha 
= X^\delta P(u)^\gamma{}_\delta P(u)^\alpha{}_\beta  
\nabla_\gamma Y^\beta\ ,
\end{equation} 
the orthogonal splitting of the covariant derivative defines a time derivative 
$\nabla_{\rm(fw)}(u) = P(u)\nabla_u$ (the spatial Fermi-Walker derivative)
and a spatial covariant derivative vector operator
$\vec\nabla(u) $ such that 
$X\cdot_u \vec\nabla(u)= \nabla(u)_{X}$.
One  then has the generalizations of the usual spatial vector analysis operators 
${\rm grad}_u f = \vec\nabla(u) f$,
${\rm curl}_u X = \vec\nabla(u) \times_u X$,
and
${\rm div}_u X = \vec\nabla(u) \cdot_u X$.
The spatially projected Lie derivative along $u$ is also needed and is given the analogous symbol $\Lie_u(u)=P(u)\Lie_u$. 
These differential operators have been studied by Bini, Carini and Jantzen.\cite{mfg}

The relative observer formulation of Maxwell's equations is well known. Projection of the differential form equations (\ref{eq:ME}) along and orthogonal to $u$ gives the spatial scalar (divergence) and spatial vector (curl) equations. Using the notation and conventions of Bini, Carini and Jantzen\cite{mfg} these are
\begin{eqnarray}
    && {\rm div}_u B(u) + H(u) \cdot_u E(u) = 0 \ , \nonumber \\
    && {\rm curl}_u E(u) - g(u) \times_u E(u)
             + [ \Lie(u)_{u} + \Theta(u) ] B(u) =0 \ ,\nonumber \\
    && {\rm div}_u E(u) - H(u) \cdot_u B(u) = 4\pi \rho(u)\ , \nonumber \\
    && {\rm curl}_u B(u) -  g(u) \times_u B(u)
             -  [ \Lie(u)_{u} + \Theta(u) ] E(u) = 4\pi J(u) \ ,
\label{eq:meq}
\end{eqnarray}
where the gravitoelectric vector field $g(u)=-a(u)=-\nabla_u \, u$ and the gravitomagnetic vector field $H(u)= 2[\dualp{u}\omega(u)^\flat]^\sharp$ 
of  the observer $u$ (sign-reversed acceleration and twice the vorticity vector field) are defined by the exterior derivative of $u$
\begin{equation}
   du^\flat =[u\wedge g(u)+ \dualp{u} H(u)]^\flat\ .
\end{equation}
The expansion scalar $\Theta (u) =-\delta u^\flat = {\rm Tr}\,\theta(u)$ appears in an additional term in the covariant derivative of $u$  as the trace of the (mixed) expansion tensor $\theta(u)$, of which the shear tensor $\sigma(u)=\theta(u)-\frac13\Theta(u)\delta$ is its tracefree part
\begin{eqnarray}
&&
  \nabla u = - a(u) \otimes u^\flat + \theta(u)-\omega(u) \ ,
\nonumber\\
&&
  u^\alpha{}_{;\beta} 
     = - a(u)^\alpha u_\beta +\theta(u)^\alpha{}_\beta -\omega(u)^\alpha{}_\beta\ .
\end{eqnarray}
The sign of the (mixed) vorticity tensor $\omega(u)$ 
depends on the index, signature and orientation conventions, and is chosen here so that its spatial dual is the commonly accepted vorticity vector; in an oriented frame one has 
\begin{equation}
\omega(u)^\alpha 
= \frac12 u_\delta \eta^{\delta\alpha\beta\gamma} 
  \omega(u)_{\beta\gamma}
= \frac12 u_\delta \eta^{\delta\alpha\beta\gamma} 
  u_{\gamma;\beta}
=[\frac12 {\rm curl}_u u]^\alpha\ .
\end{equation}

This representation of Maxwell's equations differs from the Ellis represen\-ta\-tion\cite{ellis,ellis98} only in the use of the spatially projected Lie derivative rather than the spatially projected covariant derivative along $u$ (spatial Fermi-Walker derivative). These two derivative operators are related by the following identity for a spatial vector field $X$ (orthogonal to $u$)
\begin{equation}
  [ \Lie(u)_{u} + \Theta(u) ] X
   = [\nabla_{\rm(fw)}(u) + \{-\sigma(u)+\omega(u)\} \rightcontract] X \ .
\end{equation}

Black hole spacetimes have two privileged observer families, the static observers with 4-velocity $m$ satisfying $\Theta(m)=0=\sigma(m)$, and the zero-angular-momentum observers (ZAMO's) with 4-velocity $n$ satisfying $\Theta(n)=0=\omega(n)$. For both $u=n,m$ there exist acceleration potentials (lapse functions)
\begin{equation}
 g(u)= - {\rm grad}_u \ln L(u)\ ,\  L(n)=N\ ,\ L(m)=M\ ,
\end{equation}
which allow the curl and acceleration terms in Maxwell's equations to be rewritten as a single term
\begin{equation}
  L(u)^{-1} {\rm curl}_u [ L(u) X(u) ] = {\rm curl}_u X(u) - g(u) \times_u X(u) \ .
\end{equation}
In a comoving coordinate system $\{t,x^i\}$ $(i=1,2,3)$ adapted to $u$, then the observer four-velocities and 
Lie derivatives simplify to
\begin{eqnarray}
   && n= N^{-1} [\partial/\partial t - \vec N ]\ ,\ 
      \Lie(n)_n = N^{-1} \Lie(n)_{ \partial/\partial t - \vec N }\ ,\nonumber\\
   && m= M^{-1} [\partial/\partial t ]\ ,\ 
      \Lie(m)_m = M^{-1} \Lie(m)_{ \partial/\partial t }\ ,
\end{eqnarray}
where $\vec N$ is the shift vector field.

\section{Spacetime integrals in $1+3$ form}

The integral form of Maxwell's equations in classical physics relates the flux of electric/magnetic field through a closed surface in space to the enclosed electric/magnetic charge and relates the time derivative of the flux through a loop in space to the line integral of the magnetic/electric field around the loop and the current passing through the loop. These integral equations are obtained from the four differential Maxwell equations by applying Gauss's divergence theorem to the two scalar equations and Stokes' theorem to the two vector equations. 

Developing the analogs of these equations in a curved spacetime requires an observer family to split the electromagnetic field and a slicing of the spacetime by a family of timelike hypersurfaces, which may be taken as the hypersurfaces associated with constant values of some time function $t$. The latter structure is needed to describe the curves, surfaces and spatial regions over which the integration is performed at some moment of ``time." If the observer family is taken to be the field 
$n=-N dt^\sharp$ of unit normals to the slicing (4-velocity field of the ``hypersurface observers"), then much simplification occurs, but if the desired observer family is not hypersurface orthogonal, as is the case even for rotating reference frames in Minkowski spacetime, one is forced to consider the more general case.
Over the years even this simple case has been the subject of many confusing discussions in the literature regarding Maxwell's equations\cite{cor} and even relativistically ``correct" rotating frame transformations.

In the integral formulation of Maxwell's equations, Stokes' theorem is applied to the 3-form exterior derivatives of $F$ and $\dual F$ using in the spacetime differential form version of Maxwell's equations  (\ref{eq:ME}).
Two distinct types of hypersurfaces must be used in this application in order to yield the spatial scalar and spatial vector information with respect to a time function and a test observer family. Let $\V(t)$ denote a family of spatial regions parametrized by the time $t$, filling a world tube in spacetime, with boundaries $\partial \V(t)$ which are closed 2-surfaces at time $t$ with the outward orientation. This first type of hypersurface leads to the integrated form of the spatial divergence equations projected along the unit normal $n$, even if expressed in terms of a different test observer family.

The second type of hypersurface is the world sheet of a family of 2-surfaces $\SS(t)$ between two hypersurfaces $t$ and $t+\epsilon$. The boundary of this world sheet consists of $\SS(t+\epsilon)$ and $\SS(t)$ with opposite inner orientations. Taking the limit of Stokes' theorem here as $\epsilon\to0$ leads to the time derivative integral equations, with the integral on the 2-surface composed of the family of boundaries $\partial \SS(t)$ between $t$ and $t+\epsilon$ reducing to a line integral on $\partial \SS(t)$ alone.  These will depend on the relative velocity of the 2-surface and the unit normal $n$. An alternative to this limiting operation is to use a transport theorem.

To discuss these questions uniformly, let ${\cal D}^{(p)}(t)$ denote a family of $p$-surface integration domains within the family of time slices and parametrized by the time $t$, namely a family of spatial regions $\V(t)$ ($p$=3) or a family of surfaces $\SS(t)$ ($p$=2) or a family of closed curves $C(t)$ ($p$=1). Let $\partial {\cal C}(t)$ denote the corresponding family of boundary $(p-1)$-surfaces.
Let $U$ be an evolution vector field whose 1-parameter
family of diffeomorphisms drags the family ${\cal D}^{(p)}(t)$ into itself, satisfying $\pounds_U \, t = U\leftcontract dt =1$.
Dragging along by $U$ is then equivalent to translation in the coordinate $t$ when completed to a system of comoving coordinates $\{t,x^i\}$.
Assuming that $U$ is timelike (or at least nonnull), the corresponding 4-velocity (or unit tangent) is
\begin{equation}\label{eq:Uhat}
\hat U 
= \gamma(\hat U,n) [ n + \nu(\hat U,n) ]
=\gamma(\hat U,u) [ u + \nu(\hat U,u) ]\ ,  
\end{equation}
so that 
\begin{equation}
U = N \gamma(\hat U,n)^{-1} \hat U\ ,
\end{equation}
where $N=-n\cdot U$ is the slicing lapse function.
Such evolution vector fields $U$ can differ by vectors which are tangent to the ${\cal D}^{(p)}(t)$, corresponding to an extra motion of the region or surface or curve into itself. 
If $U_t$ denotes the 1-parameter family of diffeomorphisms associated with $U$ (its ``flow"), then 
\begin{equation}
{\cal D}^{(p)}(t)=U_{t-t_0} {\cal D}^{(p)}(t_0)\ .
\end{equation}
A similar equation holds for the corresponding boundaries.

For a spacetime $p$-form $\Omega$ and $p$-surface integration domain ${\cal D}^{(p)}(t)$ within the hypersurface of time $t$, the transport theorem is then
\begin{equation} \label{trasporto}
\frac{d}{dt} \int_{{\cal D}^{(p)}(t)}\Omega 
= \int_{{\cal D}^{(p)}(t)}\Lie_U\Omega \ ,
\end{equation}
where
\begin{equation}
  \Lie_U\Omega =  U\leftcontract d \Omega + d [U\leftcontract \Omega]
\end{equation}
is a useful formula for the Lie derivative of any $p$-form, and which allows the second term to be integrated to the boundary using Stokes' theorem
\begin{equation}
  \int_{{\cal D}^{(p)}(t)} d\zeta 
= \int_{\partial{\cal D}^{(p)}(t)}\zeta
\end{equation}
where $\zeta$ is a $(p-1)$-form.

The differential form integrals require no metric, but they can be converted to more familiar but equivalent metric notation for volume, surface and line integrals, which is related to the $p$-form integral notation for a function $f$ and a spatial vector field $X(u)$ by
\begin{eqnarray}
\qquad
\int_{\V(t)} \, f \,\eta(u) &=& \int_{\V(t)} \, f \,dV(u) \ ,
\nonumber\\  
\qquad 
\int_{\SS(t)} \,\dualp{u} X(u)^\flat &=& \int_{\SS(t)} X(u) \cdot_u dS(u) \ ,
\nonumber\\  
\qquad 
\int_{C(t)}  X(u)^\flat 
&=& \int_{C(t)} X(u)\cdot_u d\ell(u)\ .
\end{eqnarray}
For example, Stokes' theorem for $p=2,3$ then becomes
\begin{eqnarray}
\int_{\SS(t)} {\rm curl}_u\, X(u) \cdot_u dS(u)
&=&
\int_{\partial \SS(t)} \, X(u) \cdot_u d\ell(u)\ ,
\nonumber\\  
\int_{\V(t)} \, {\rm div}_u\, X(u) \, dV(u) 
&=&
\int_{\partial\V(t)} \, X(u) \cdot_u dS(u)  \ . 
\end{eqnarray}

The transport theorem is the result of the calculation
\begin{eqnarray}\label{eq:trasporto}
 \left.\frac{d}{dt}\right\vert_{t=t_0} \int_{{\cal D}^{(p)}(t)} 
         \, \Omega
&=&\left.\frac{d}{dt}\right\vert_{t=t_0} \int_{U_{t-t_0}{\cal D}^{(p)}(t_0)} 
         \, \Omega
\nonumber\\ 
&=&\left.\frac{d}{dt}\right\vert_{t=t_0} \int_{{\cal D}^{(p)}(t_0)} 
         \, U_{t-t_0}^{-1}\Omega
=\int_{{\cal D}^{(p)}(t_0)} \, \pounds_U \Omega \ ,
\end{eqnarray}
transforming the integral and using the definition of the Lie derivative.
This single transport theorem corresponds to the line, surface and volume integral formulas found in some older textbooks on electromagnetism and collected together and expressed 
in conventional notation in appendix D of van Bladel\cite{vanbladel} and discussed in the appendix.

\section{Integral $1+3$ form of Maxwell's equations}

The ``spatial scalar" Maxwell equations in integral form come from integrating Maxwell's equations (\ref{eq:ME})
over a 3-surface region $\V(t)$ of a time coordinate hypersurface.
Integrating  first $dF =0$ leads to
$$ 
0=\int_{\V(t)} dF 
=\int_{\partial \V(t)} F
=\int_{\partial \V(t)}[u\wedge E(u)+\dualp{u}B(u)]^\flat\ .
$$
However, restricting the 1-form $n^\flat$
to a constant time coordinate hypersurface implies
\begin{equation}\label{eq:Unu}
0=-Ndt|_{t=t_0}=n^\flat|_{t=t_0}
=\gamma(n,u)[u^\flat+\nu^\flat(n,u)]|_{t=t_0}
\rightarrow
u^\flat |_{t=t_0}=-\nu^\flat(n,u) |_{t=t_0}
\end{equation} 
so the right hand side of the 
integral equation becomes
$$
\int_{\partial \V(t)}\dualp{u}[-{\nu}(n,u)\times_ u {E}(u)+{B}(u)]^\flat\ .
$$
We then have

{\bf$\bullet$ Gauss's law for the magnetic field}
\begin{eqnarray} 
\nonumber\\
&\int_{\partial \V(t)} [{B} (u)-{\nu}(n,u)\times_ u {E} (u)] \cdot_ u dS (u) =0\ .
\end{eqnarray}
The final expression uses the more conventional metric notation for volume, surface and line integrals.

Analogously,  the integral Maxwell equation dual to this one comes from integrating the dual Maxwell equation.
The right hand side of this integral is
\begin{eqnarray}
 -4\pi\int_{\V(t)} \dual J^\flat
&=& -4\pi\int_{\V(t)} \dual [\rho(u) u + J(u)]^\flat
\nonumber \\
&=&4\pi\int_{\V(t)} \dualp{u}[\rho(u)- {\nu}(n,u)\cdot_ u {J}(u)]
\nonumber \\
&=&4\pi\int_{\V(t)} [\rho(u)-{\nu}(n,u)\cdot_ u {J} (u)] \, dV (u)\ , \nonumber 
\end{eqnarray}
so that one has instead (the spacetime duality operation on $F$ is equivalent to $[E(u),B(u)] \mapsto  [-B(u),E(u)]$)
\begin{itemize}
\item {\bf Gauss's law for the electric field (Coulomb's law)}
\end{itemize}
\begin{eqnarray}
\int_{\partial \V(t)} [{E} (u)+{\nu}(n,u)\times_ u {B} (u)] \cdot_ u dS(u)
\nonumber\\ \qquad 
= 4\pi\int_{\V(t)}  [\rho(u)-{\nu}(n,u)\cdot_ u {J} (u)] \, dV (u)\ .
\end{eqnarray}
The extra velocity terms in both Gauss law equations correct for the motion of the observers with respect to the time hypersurface in which the integration takes place.

The ``spatial vector" Maxwell equations in integral form come from applying the transport theorem (\ref{trasporto}) with $\Omega=F$ and 
$\Omega=\dual F$ respectively on a 2-surface $\SS(t)$.
In the first case,  the left hand side of this equation is
\begin{eqnarray*}
\frac{d}{dt} \int_{\SS(t)} F 
&=&\frac{d}{dt} \int_{\SS(t)} [u\wedge E(u) + \dualp{u} B(u)]^\flat 
\nonumber\\
&=&\frac{d}{dt} \int_{\SS(t)} \dualp{u}[-{\nu}(n,u)\times_ u {E}(u) + {B}(u)]^\flat \nonumber \\
&=&\frac{d}{dt} \int_{\SS(t)} [{B} (u)- {\nu}(n,u)\times_ u {E} (u)] 
                              \cdot_ u \, dS (u)\ , 
\end{eqnarray*}
while the right  hand side is
\begin{eqnarray*}
\int_{\SS(t)} \Lie_U F 
=\int_{\SS(t)} [U\leftcontract dF + d(U\leftcontract F)]
=\int_{\partial \SS(t)} U\leftcontract F \ ,
\end{eqnarray*}
where Maxwell's first equation and Stokes' theorem have been used.
Using (\ref{eq:FEB}) and (\ref{eq:Uhat}), this 1-form evaluates to
\begin{eqnarray*}
U\leftcontract F 
&=& N \gamma(\hat U,n)^{-1} \hat U\leftcontract F 
= -N \gamma(\hat U , n)^{-1} E(\hat U)\nonumber \\
&
=&-N \gamma(\hat U,n)^{-1} \gamma (\hat U ,u)  P(u,\hat U)^{-1}
[ E(u)+\nu (\hat U,u)\times_u B(u)] \nonumber \\
&
=&-N \gamma (\hat U ,n)^{-1} \gamma (\hat U ,u)
[E(u)+\nu (\hat U,u)\times_u B(u) +u [\nu(\hat U, u)\cdot E(u)]]^\flat \nonumber 
\end{eqnarray*}
so that
\begin{eqnarray*}
&
U\leftcontract F |_{t=t_0}
=
-N \gamma (\hat U ,n)^{-1} \gamma (\hat U ,u)
[E(u)+\nu (\hat U,u)\times_u B(u) 
\\ &\qquad
-\nu(n,u)[\nu(\hat U, u)\cdot E(u)]]^\flat\ \vert_{t=t_0}\ ,
\end{eqnarray*} 
where the ``inverse" of the mixed projector $P(u,\hat U)=P(u)P(\hat U)$ and 
the electric field transformation law have been  given by Jantzen, Carini and Bini\cite{mfg} (equations (4.7) and (4.14))  and equation (\ref{eq:Unu}) has been used.
The final form is then
\begin{itemize}
\item {\bf Faraday's law} 
\end{itemize}
\begin{eqnarray}
&&
 -\frac{d}{dt} \int_{\SS(t)} [{B} (u)- {\nu}(n,u)\times_ u {E} (u)] \cdot_ u dS (u)  \nonumber \\
&&\qquad
 =\int_{\partial S(t)} 
N \gamma (\hat U ,n)^{-1} \gamma (\hat U ,u)
[E(u)+\nu (\hat U,u)\times_u B(u) \nonumber \\
&&\qquad\qquad\qquad
-\nu(n,u)[\nu(\hat U, u)\cdot E(u)]]\cdot_ u d\ell (u) \ ,
\end{eqnarray}
where the right hand side merely represents the line integral of the component of the Lorentz force along the direction of the curve, namely
\begin{equation}
 \int N [E(n)+\nu (\hat U,n)\times_n B(n)] \cdot_n d\ell(n) \ ,
\end{equation}
corrected for the time $t$ rather than the proper time associated with $n$ and then re-expressed in terms of the generic observer $u$. 
In fact one can write this right hand side in the alternative form
\begin{eqnarray}
\qquad
 \int_{\partial S(t)} 
L(u) 
[E(u)+ {\cal U}\times_u \{ B(u) - \nu(n,u)\times_u E(u) \} ]
\cdot_ u d\ell (u) \ ,
\end{eqnarray}
where the ``lapse-like function" $L(u)=N \gamma (u ,n)^{-1} $ reparametrizes the $u$ proper time derivative to the $t$ time derivative in this equation and
\begin{equation}
{\cal U} = \frac{\gamma(u,n)}{\gamma(\hat U,n)} P(u) \hat U 
=  \frac{\gamma(\hat U,u) \gamma(u,n)}{\gamma(\hat U,n)}  \nu(\hat U,u)
\end{equation}
is the 4-velocity of the curve $\partial \SS(t)$ reparametrized by the $u$ proper time and projected into the local rest space of $u$, which would have spatial coordinate components ${\cal U}^i=dx^i /d\tau_u$ in comoving coordinates adapted to $u$ (denoted by $\undersim{v}$ in Thorne and Macdonald\cite{thomac}). The extra term containing it corrects for the motion of the curve with respect to the observer $u$.

The dual equation is obtained in a similar way, with the additional term resulting from Maxwell's second equation
\begin{eqnarray*}
&&\int_{\SS(t)} U\leftcontract d\dual F  
  =-4\pi\int_{\SS(t)} U\leftcontract \dual J^\flat \nonumber \\
&& =-4\pi\int_{\SS(t)}  
N\gamma (\hat U, n)^{-1} \gamma (\hat U, u)[-\rho (u) \nu (\hat U, u)+J(u) \nonumber \\
&& \qquad
+\nu(n,u) \times_u [\nu(\hat U,u)\times_u J(u)])]
  \cdot_ u dS (u) \ ,
\end{eqnarray*}
where the relations 
\begin{equation}\label{eq:UcJ}
U \leftcontract \dual J^\flat =\frac{N}{\gamma (\hat U, n)}\hat U \leftcontract \dual J^\flat =
\frac{N}{\gamma (\hat U, n)}  \dualp {\hat U} J(\hat U)^\flat
\end{equation}
and
$$\eta (\hat U)=\gamma (\hat U ,u)[\eta (u) -u\wedge \dualp u \nu (\hat U ,u)]$$
have been used.
This term can also be rewritten as above
\begin{equation}
\int_{\SS(t)} L(u)[J(u) - {\cal U}\{ \rho (u)-\nu(n,u)\cdot_u J(u)\}] \cdot_u d S(u) \ ,
\end{equation}
where again the ${\cal U}$ term corrects for the motion of the surface as above.

The final form in the original notation is then
\begin{itemize}
\item {\bf Ampere's law}
\end{itemize}
\begin{eqnarray}
\qquad
&&-\frac{d}{dt} \int_{\SS(t)} 
  [{E} (u)+ {\nu}(n,u)\times_ u {B} (u)] \cdot_ u dS (u)  
\nonumber \\
\qquad
&& = -\int_{\partial \SS(t)} 
N \gamma (\hat U ,n)^{-1} \gamma (\hat U ,u)
[B(u)-\nu (\hat U,u)\times_u E(u) 
\nonumber\\ &&\qquad
-\nu(n,u)[\nu(\hat U, u)\cdot B(u)]]\cdot_ u d{l} (u) 
\nonumber \\
\qquad
&& \quad +4\pi\int_{\SS(t)}
N\gamma (\hat U, n)^{-1} \gamma (\hat U, u)[-\rho (u) \nu (\hat U, u)+J(u) 
\nonumber \\
\qquad
&&
\qquad +\nu(n,u)
\times_u [\nu(\hat U,u)\times_u J(u)]]
  \cdot_ u dS (u) \ .
\end{eqnarray}

The integral form of charge conservation is obtained by applying the $p=3$ version of the theorem (\ref{eq:trasporto}),
integrating $\Omega=\dual J^\flat$ on a hypersurface region $\V(t)$, using $ d \dual J^\flat = 0$ to eliminate the first term 
in the Lie derivative and then applying Stokes' theorem to the second term $d U\leftcontract \dual J^\flat$, and finally
using (\ref{eq:UcJ}) to re-express $U\leftcontract \dual J^\flat$. Alternatively one can use Coulomb's law for the electric field to replace the flux integral by the charge integral in the preceding equation, where $\SS(t) =\partial \V(t)$ so that $\partial\SS(t)$ is the empty set. The result is
\begin{itemize}
\item {\bf Charge conservation}
\end{itemize}
\begin{eqnarray}
\qquad
&&\frac{d}{dt} \int_{\V(t)} 
  [\rho(u) - \nu(n,u)\cdot_u J(u)] dV (u)  
\nonumber \\
\qquad
&& =\int_{\partial\V(t)}
N\gamma (\hat U, n)^{-1} \gamma (\hat U, u)[-\rho (u) \nu (\hat U, u)+J(u) 
\nonumber \\
\qquad
&&
\qquad +\nu(n,u)
\times_u [\nu(\hat U,u)\times_u J(u)]]
  \cdot_ u dS (u) \ .
\end{eqnarray}

It is important to note that each of the integrals which appear in the integral form of Maxwell's equations are merely expressed in terms of the observer with 4-velocity $u$, but the values of these integrals are independent of this observer, and reduce to the values assumed when the observer family is the family of normal observers ($u=n$ so $\nu(n,u)=0$). These values are the Thorne and Macdonald equations\cite{thomac}
\begin{eqnarray}
&& \int_{\partial \V(t)} {B} (n) \cdot_ n  dS (n) =0\ , 
\nonumber\\
&& \int_{\partial \V(t)} {E} (n)\cdot_ n  dS (n) 
= 4\pi\int_{\V(t)}  \rho(n) \, dV (n) \ , 
\nonumber\\
&& \int_{\partial \SS(t)} N [{E} (n)+ \nu(\hat U, n)\times_ n {B}(n)]
 \cdot_ n d\ell(n)
=-\frac{d}{dt} \int_{\SS(t)} {B} (n) \cdot_ n dS (n) \ , 
\nonumber\\
&&\int_{\partial \SS(t)} N [{B} (n)-\nu(\hat U, n)\times_ n {E}(n)]
\cdot_ n d\ell(n)
     \nonumber \\
&&=\frac{d}{dt} \int_{\SS(t)} {E} (n) \cdot_ n dS (n) + 
4\pi\int_{\SS(t)} N [{J}(n)-\nu(\hat U, n) \rho(n)] \cdot_ n dS (n)\ ,  
\end{eqnarray}
and if the surface $\SS(t)$ is also at rest with respect to these observers ($\nu(\hat U, n)=0$), the latter two simplify further.
If the surface $\SS(t)$ is at rest with respect to the observers $u$, then $\hat U = u$ and $\nu(\hat U, u)=0$, leading to a different simplication.

\section{Lines of force}

In Minkowski spacetime, it is customary to picture electric and magnetic fields at some moment of time by drawing their corresponding lines of force on space or some cross-sectional plane surface within space. The density of field lines (in charge-free regions for the electric field) can be taken to directly reflect the magnitude of the field strength since the integrated Maxwell divergence equations show that the flux is conserved, i.e., in any flux tube whose lateral sides (by definition) consist of field lines, the number of field lines through the ends of the flux tube is the same at both ends. Field lines are not created or destroyed. Thus the number of field lines passing through a fixed area can be taken as proportional to the field intensity. For a closed surface, the total electric flux directly determines the enclosed charge, while the total magnetic flux is zero.

The same is true for the hypersurface observers in a general spacetime, since the same integral form of these Maxwell equations holds. The flux of electric and magnetic field lines is conserved and the integral of the charge density over a volume determines the enclosed charge, which is proportional to the electric flux out of its boundary, while the total magnetic flux out of a volume is always zero. The electric lines of force of a point charge at rest near a black hole and the magnetic lines of force around a black hole immersed in an asymptotically uniform magnetic field, as seen by the ZAMO hypersurface observers, were first introduced by Hanni and Ruffini\cite{hanni1} and used by Ruffini and collaborators\cite{hanni2,ruf1,ruf2} to study electrodynamic properties of black holes 
(see also later work by Thorne et al\cite{thoprimac}).

However, for general observers, the separate fluxes of the electric and magnetic fields are no longer conserved and one loses the nice connection between field intensity and the density of flux lines. Similarly the classic relation between electric flux and charge is lost, and the total magnetic flux out of a closed surface is in general not zero. Even the integral of the charge density as observed by $u$ over a volume does not directly determine the enclosed charge. Instead one is forced to work with the combinations of the fields which are equivalent to transforming back to the hypersurface observer quantities, which is why the integral form of Maxwell's equations is so complicated for a general observer. However, van Bladel\cite{vanbladel} has shown that in spite of this complication, it is still possible to use them in describing certain kinds of problems in accelerated frames of reference. 

\nonumsection{Acknowledgements}
\noindent
The authors thank Remo Ruffini for useful discussion and support.

\nonumsection{References}


\newpage 
\appendix{. Time derivative integral formulas: transport theorems}

The elegant differential forms statement of Stokes' theorem simultaneously captures the vector analysis integral theorems for line integrals, surface integrals and volume integrals and their boundaries on a 3-dimensional Riemannian manifold. In a similar way the modern statement of the transport theorem contains the individual results in vector analysis language. First the flat spacetime versions of these are recovered, and then the hypersurface observer versions are obtained.

\subsection{Van Bladel integral formulas}

Some older textbooks on electrodynamics contain these transport theorems, restated by Van Bladel\cite{vanbladel}. Starting from the the general form (\ref{eq:trasporto}) of the transport theorem,
specialize to 
flat spacetime and Minkowski coordinates $\{t,x^i\}$, with $n=\partial_t$ usual unit time direction vector field, $X(n)=X(n)^i \partial_i$ a spatial vector field, $f$ a spacetime scalar, and the notation $d\ell(n)= dx^i \partial_i$, $dS(n)=\delta^{ij} \epsilon_{jkl} dx^k \wedge dx^l \partial_i $,
$dV(n)= dx^1\wedge dx^2 \wedge dx^3$. One finds
\begin{eqnarray}
\label{vanbla1}
&&
 \frac{d}{dt} \int_{C(t)} X(n)\cdot_n d\ell(n)
\\ &&
  = \int_{C(t)} 
\left[ \frac{\partial X(n)}{\partial t}
        +{\rm grad}\, [ X(n)\cdot \nu(n) ] 
        + [{\rm curl}\,  X(n)]\times \nu(n) 
\right] \cdot_n d\ell(n)\ ,\nonumber\\
\label{vanbla2}
&&
 \frac{d}{dt} \int_{{\cal S}(t)} X(n)\cdot_n dS(n)
\\ &&
  = \int_{{\cal S}(t)} 
\left[ \frac{\partial X(n)}{\partial t}
        +\nu(n) {\rm div}\, X(n) 
        + {\rm curl}\, [ X(n)\times \nu(n) ]
\right] \cdot_n dS(n)\ ,\nonumber\\
\label{vanbla3}
&&
 \frac{d}{dt} \int_{{\cal V}(t)} f(n) \, dV(n)
\\ &&
  = \int_{{\cal V}(t)} 
\left[ \frac{\partial f(n)}{\partial t}
        +f(n) {\rm div}\, \nu(n) 
        + \nu(n) \cdot {\rm grad}\, f(n)
\right] dV(n)\ .\nonumber
\end{eqnarray}
where
\begin{equation}
  \frac{\partial f}{\partial t}
+ \nu(n) \cdot {\rm grad}\, f
 =\frac{D f}{d t}
\end{equation}
is the total derivative along the curve parametrized by coordinate time $t$ and having the tangent vector $U=\partial_t +\nu(n)^i\partial_i$.

For example, consider deriving the second of these. Write the curl term in equation (\ref{vanbla2}) as
\begin{equation}
\label{curl}
{\rm curl} [X(n)\times \nu(n) ]^a+\nu(n)^a [{\rm div}\, X(n)] = \Lie_{\nu(n)} X(n)^a +X(n)^a [{\rm div}\, \nu(n)] \ .
\end{equation}
Taking into account the relation 
$\Lie_{\partial_t} X(n)^a=\partial_t X(n)^a$ 
and by adding this term to both sides of equation (\ref{curl}) one finds 
\begin{eqnarray}
\label{eq2}
&\partial_t X(n)^a + {\rm curl} [X(n)\times \nu(n) ]^a 
    + \nu(n)^a [{\rm div}\, X(n)] \nonumber \\
&= \Lie_{U} X(n)^a + X(n)^a [{\rm div}\, \nu(n)] \ .
\end{eqnarray}
Finally, contraction by $dS(n)_a=\frac12\eta(n)_{abc}dx^b\wedge dx^c$ gives equation (\ref{vanbla2})
\begin{eqnarray}
\label{eq2bis}
&\left[\partial_t X(n)^a+{\rm curl} [X(n)\times \nu(n) ]^a+\nu(n)^a [{\rm div}\, X(n)]\right]dS(n)_a =\nonumber \\
&= \left[\Lie_{U} X(n)^a +X(n)^a [{\rm div}\, \nu(n)]\right]\frac12\eta(n)_{abc} dx^b\wedge dx^c\nonumber \\
&= {}^{*(n)}[\Lie_U [\eta(n)\rightcontract X(n)]]_{bc}\frac12 dx^b\wedge dx^c \ ,
\end{eqnarray}
for which $\Omega^{(2)}={}^{*(n)}X(n)$.

Analogously equation (\ref{vanbla3}) corresponds to $\Omega=f(n) \eta(n)$. In fact one has
\begin{equation}
\Lie_U [f(n)]= \partial_t f(n) +\nu(n) \cdot \nabla f(n)\ ,
\end{equation}
so that
\begin{equation}
{}^{*(n)}\Lie_U [f(n)\eta(n)]= [\partial_t f(n) +\nu(n) \cdot \nabla f(n)+ ({\rm div}\,\nu (n) )\, f(n)]\ ,
\end{equation}
where the relation ${}^{*(n)}\Lie_U [\eta(n)]={\rm div}\,\nu (n)$ has been used.

The first relation (\ref{vanbla1}) corresponds to
$\Omega^{(1)}=X(n)^\flat$.
This is easily obtained starting from the vector identity
\begin{equation}
\left[ [{\rm curl}\,  X(n)]\times \nu(n) \right]_a 
= -\nu(n)^c \nabla_a X(n)_c + \nu(n)^c \nabla_c X(n)_a
\end{equation}
and the Lie derivative relation
\begin{equation}
\nu(n)^c \nabla_c X(n)_a= \Lie_{\nu(n)} X(n)_a - X(n)_b \nabla_a \nu (n)^b\ ,
\end{equation}
so that
\begin{equation}
\partial_t X_a + [({\rm curl}\,  X(n))\times \nu(n)]_a 
= \Lie_U X(n)_a -\nabla_a (X(n)\cdot \nu(n)) \ .
\end{equation}

\subsection{Thorne and MacDonald}

The corresponding generalizations to any hypersurface observers in a general spacetime, where $n$ is the unit normal to a slicing by spacelike hypersurfaces, are the Thorne-MacDonald equations (2.27), (2.26), and (2.25) respectively \cite{thomac}
\begin{eqnarray}
&&
 \frac{d}{dt} \int_{C(t)} X(n)\cdot_n d\ell(n)
\\ &&
  = \int_{C(t)} N
\left[\strut \nabla(n)_n  X(n)
        + \theta\rightcontract X(n)
        + [ \nabla(n) \times_n X(n) ] \times_n \nu(n)
\right] \cdot_n d\ell(n)\ ,\nonumber\\
&&
 \frac{d}{dt} \int_{{\cal S}(t)} X(n)\cdot_n dS(n)
\\ &&
  = \int_{{\cal S}(t)} N
\left[\strut \nabla(n)_n X(n)
        +  [ \Theta(n) -\theta\rightcontract ] X(n) 
+\nu(n) [ \nabla(n)\cdot_n X(n) ]
\right.\nonumber\\ 
&& \left.\strut\qquad
        + \nabla(n) \times_n [ N X(n)\times \nu(n) ]
\right] \cdot_n dS(n)\ ,\nonumber\\
&&
 \frac{d}{dt} \int_{{\cal V}(t)} f(n)\cdot_n dV(n)
\\ &&
  = \int_{{\cal V}(t)} 
\left[\strut N [\nabla(n)_n + \Theta(n)] f(n)
        +\nabla(n) \cdot [ N f(n) \nu(n) ]
\right] dV(n)\ .\nonumber
\end{eqnarray}

Of course these equations reduce to Van Bladel's when $N=1$, $\theta=0=\Theta (n)$. The precises Thorne-MacDonald form of the last two of these equations is obtained by using Stokes' theorem on the final curl and divergence terms to convert them respectively to a line integral and a surface integral.

\end{document}